\documentclass[12pt]{article}
\usepackage{enumerate}
\usepackage{amsfonts}
\usepackage{amssymb}

\newcounter{defin}  \newcounter{lemma}  \newcounter{theorem}
\newcounter{property} \newcounter{corol}  \newcounter{remark} \newcounter{example}

\newenvironment{lemma}{\par\refstepcounter{lemma}
     \textbf{Lemma \thelemma.} }{\rm\par}
\newenvironment{theorem}{\par\refstepcounter{theorem}
     \textbf{Theorem \thetheorem.}\ }{\rm\par}
\newenvironment{property}{\par\refstepcounter{property}
     \textbf{Proposition \theproperty.}\ }{\rm\par}
\newenvironment{corollary}{\par\refstepcounter{corol}
     \textbf{Corollary \thecorol.} }{\rm\par}

\newenvironment{remark}{\par\refstepcounter{remark}
     \textbf{Remark \theremark.}}{\rm\par}

\begin{document}
\title{The Convex Closure of the Output Entropy of Infinite Dimensional
Channels and the Additivity Problem}
\author{M.E.Shirokov \thanks{Steklov Mathematical Institute, Moscow,
Russia, e-mail:msh@mi.ras.ru.}}
\date{}
\maketitle

\section{Introduction}

The central role in quantum information theory is played by the
notion of a quantum channel, transforming states of one (input)
quantum system into states of other (output) quantum system. The
important characteristics of a quantum channel are the output
entropy and its convex closure\footnote{It is also used the term
"lower envelope" \cite{B&R}.} (cf.\cite{J&T}), which is widely
used in quantum information theory, sometimes, implicitly. For
example, the Holevo capacity of a quantum channel with finite
output entropy can be defined as the maximal difference between
the output entropy and its convex closure. Another example is the
notion of entanglement of formation (\textup{EoF}) of a state in
bipartite system. Indeed, it is easy to see that the definition of
the \textup{EoF} in the finite dimensional case  implies it
coincidence with the convex closure of the output entropy of a
partial trace \cite{B&Ko},\cite{A&B}. In the infinite dimensional
case the \textup{EoF} can be directly defined as the convex
closure of the output entropy of a partial trace. The advantages
of this definition and its relations with some others are
discussed in \cite{Sh-3}.

The representation of the convex closure of the output entropy (in
what follows, \textup{CCoOE}) of an arbitrary infinite dimensional
channel and some its properties are obtained in \cite{Sh-3}. In
this paper we develop these results and show their applications to
the additivity problem.

In section 3 the continuity conditions for the \textup{CCoOE} are
considered (propositions \ref{hf-p-1} and \ref{hf-p-2}, corollary
\ref{hf-c-7}). The nontrivial example of channel with continuous
and bounded \textup{CCoOE} is presented.

In section 4 the superadditivity property of the \textup{CCoOE} is
considered. For partial trace channels these property means strong
superadditivity of the EoF \cite{A&B}. In the finite dimensional
case superadditivity of the \textup{CCoOE} for given two channels
is equivalent to the additivity of the Holevo capacity for these
channels with arbitrary constraints \cite{H-Sh-1}. Moreover, the
global additivity conjecture can be expressed as the
superadditivity of the \textup{CCoOE} for partial trace channels
(= superadditivity of the EoF) \cite{Shor-e-a-q}. In the infinite
dimensional case the relations between different addivity
properties are more complex. The main problem in this case
consists in existence of pure states with infinite entropy of
partial traces, which can be called superentangled (see remark 4
in \cite{Sh-2}). It is this problem that up to now prevented to
prove the \textup{CCoOE}-analog of theorem 3 in \cite{Sh-2},
stated that the additivity of the Holevo capacity for all finite
dimensional channels implies the additivity of the Holevo capacity
for all infinite dimensional channels with arbitrary constraints
and to show the superadditivity of the \textup{CCoOE} (and even
additivity of the minimal output entropy) for two infinite
dimensional channels with one of them a direct sum of noiseless
and entanglement-breaking channels despite the fact that
additivity of the Holevo capacity for these channels with
arbitrary constraints is derived in \cite{Sh-2} from the
corresponding finite dimensional results
\cite{Shor-e-b-c},\cite{H-Sh-1}. In this paper we overcome this
difficulties by using special approximation result (lemma
\ref{hf-l-1}), based on the continuity property of the
\textup{CCoOE} and some other observations from \cite{Sh-3}.

The main result of this paper is the statement that the
superadditivity of the \textup{CCoOE} for all finite dimensional
channels implies the superadditivity of the \textup{CCoOE} for all
infinite dimensional channels (theorem \ref{hf-th-1} and corollary
\ref{hf-c-10}), which implies the analogous statements for the
strong superadditivity of the EoF (corollary \ref{hf-c-11}) and
for the additivity of the minimal output entropy (corollary
\ref{hf-c-12}). This result and theorem 3 in \cite{Sh-2} provide
infinite dimensional generalization of Shor's theorem
\cite{Shor-e-a-q}, stated equivalence of different additivity
conjectures (theorem \ref{hf-th-2}). The approximation technic
used in the proof of the above result also makes possible to
derive the superadditivity of the \textup{CCoOE} (and hence the
additivity of the minimal output entropy) for two infinite
dimensional channels with one of them a direct sum of noiseless
and entanglement-breaking channels (proposition \ref{hf-p-3}) from
the additivity of the Holevo capacity for these channels with
arbitrary constraints. Some consequences of this result are
considered (corollary \ref{hf-c-1} and remark \ref{hf-r-3}).

The role of the superadditivity of the \textup{CCoOE} is stressed
by the observation in \cite{H-comp-ch},\cite{R-comp-ch} that
validity of this property for some pair of channels means its
validity for the pair of complementary channels. By theorem 1 in
\cite{H-Sh-1} in the finite dimensional case this result can be
reformulated in terms of the additivity of the Holevo capacity. In
the infinite dimensional case the situation is more difficult, but
some conditional result in this direction can be proved
(proposition \ref{hf-p-4}). This and the above observations leads
to extension of the class of infinite dimensional channels for
which the superadditivity of the \textup{CCoOE} and the additivity
of the Holevo capacity with arbitrary constraints are proved
(corollary \ref{hf-c-2}).

In the Appendix 5.1 some general continuity condition for the
quantum entropy applicable to noncompact and nonconvex sets of
states is considered (proposition \ref{hf-cont-cond}, corollaries
\ref{hf-cont-cond-1} and \ref{hf-cont-cond-2}).

\section{Preliminaries}

Let $\mathcal{H}$ be a separable Hilbert space,
$\mathfrak{B}(\mathcal{H})$ be the set of all bounded operators on
$\mathcal{H}$, $\mathfrak{T}( \mathcal{H})$ be the Banach space of
all trace-class operators with the trace norm $\Vert\cdot\Vert_{1}$
and $\mathfrak{S}(\mathcal{H})$ be the closed convex subset of
$\mathfrak{T}(\mathcal{H})$ consisting of all density operators on
$\mathcal{H}$, which is complete separable metric space with the
metric defined by the trace norm. Each density operator uniquely
defines a normal state on $\mathfrak{B}(\mathcal{H})$ \cite{B&R},
so, in what follows we will also for brevity use the term "state".

We denote by $\mathrm{co}\mathcal{A}$ ($\overline{\mathrm{co}}
\mathcal{A}$) the convex hull (closure) of a set $\mathcal{A}$ and
by $\mathrm{co}f$ ($\overline{\mathrm{co}}f$) the convex hull
(closure) of a function $f$ \cite{J&T}. We denote by
$\mathrm{ext}\mathcal{A}$ the set of all extreme points of a
convex set $\mathcal{A}$.

Let $\mathcal{P}$ be the set of all Borel probability measures on
$\mathfrak{S}(\mathcal{H})$ endowed with the topology of weak
convergence \cite{Bil},\cite{Par}.  Since
$\mathfrak{S}(\mathcal{H})$ is a complete separable metric space
$\mathcal{P}$ is a complete separable metric space as well
\cite{Par}. Let $\widehat{\mathcal{P}}$ be the closed subset of
$\mathcal{P}$ consisting of all measures supported by the closed
set $\mathrm{ext}\mathfrak{S}(\mathcal{H})$ of all pure states.

The \textit{barycenter} of the measure $\mu$ is the state defined by
the Bochner integral
\[
\bar{\rho}(\mu)=\int\limits_{\mathfrak{S}(\mathcal{H})}\sigma
\mu(d\sigma).
\]

For arbitrary state $\rho$ in $\mathfrak{S}(\mathcal{H})$ let
$\mathcal{P}_{\{\rho\}}$ (corresp.
$\widehat{\mathcal{P}}_{\{\rho\}}$) be the subset of $\mathcal{P}$
(corresp. $\widehat{\mathcal{P}}$) consisting of measures with the
barycenter $\rho$.

A collection of states $\{\rho_{i}\}$ with corresponding
probability distribution $\{\pi_{i}\}$ is conventionally called
\textit{ensemble} and is denoted by $\{\pi _{i},\rho _{i}\}$.  In
this paper we will consider ensemble of states as a partial case
of probability measure, so that notation
$\{\pi_{i},\rho_{i}\}\in\mathcal{P}_{\{\rho\}}$ means that
$\rho=\sum_{i}\pi_{i}\rho_{i}$. An ensemble consisting of finite
number of states is denoted by $\{\pi_{i},\rho_{i}\}^{\mathrm{f}}$
and is also called a measure with finite support.

We will use the following extension of the von Neumann entropy
$S(\rho)=-\mathrm{Tr}\rho\log\rho$ of a state $\rho$ to the set of
all positive trace class operators (cf.\cite{L})
$$
H(A)=-\mathrm{Tr}A\log
A+\mathrm{Tr}A\log\mathrm{Tr}A=(\mathrm{Tr}A)
S(A/\mathrm{Tr}A),\quad \forall A
\in\mathfrak{T}_{+}(\mathcal{H}).
$$
Nonnegativity, concavity and lower semicontinuity of the von Neumann
entropy $S$ on the set $\mathfrak{S}(\mathcal{H})$ imply the same
properties of the entropy $H$ on the set
$\mathfrak{T}_{+}(\mathcal{H})$.

Let $\mathcal{H},\mathcal{H}^{\prime }$ be a pair of separable
Hilbert spaces which we shall call correspondingly input and
output space. A channel $\Phi $ is a linear positive trace
preserving map from $\mathfrak{T}(\mathcal{ H })$ to
$\mathfrak{T}(\mathcal{H}^{\prime })$ such that the dual map $\Phi
^{\ast }:\mathfrak{B}(\mathcal{H}^{\prime
})\mapsto\mathfrak{B}(\mathcal{H})$ is completely positive.

The important characteristic of a quantum channel $\Phi$ is the
output entropy $H_{\Phi}(\rho)=H(\Phi(\rho))$ - concave lower
semicontinuous function on the input state space
$\mathfrak{S}(\mathcal{H})$. It is shown in \cite{Sh-3} that the
convex closure of the output entropy (\textup{CCoOE}) of an
arbitrary quantum channel $\Phi$ is defined by the expression
\begin{equation}\label{H-fun-rep-g}
\overline{\mathrm{co}}H_{\Phi}(\rho)=
\inf_{\mu\in\mathcal{P}_{\{\rho\}}}\int\limits_{\mathfrak{S}(\mathcal{H})}H_{\Phi}(\sigma)
\mu(d\sigma)=\inf_{\mu\in\widehat{\mathcal{P}}_{\{\rho\}}}
\int\limits_{\mathrm{ext}\mathfrak{S}(\mathcal{H})}H_{\Phi}(\sigma)
\mu(d\sigma)
\end{equation}
for all $\rho$ in $\mathfrak{S}(\mathcal{H})$ and that the last
infimum in this expression is always achieved at some measure
$\mu_{\rho}^{\Phi}$ in $\widehat{\mathcal{P}}_{\{\rho\}}$.

\begin{remark}\label{hf-r-1}
There exist channels $\Phi$ and states $\rho$ such that each
optimal measure $\mu_{\rho}^{\Phi}$ is purely nonatomic. Indeed,
let $\Phi_{0}$ be the partial trace channel and $\rho_{0}$ be the
separable state constructed in \cite{H-Sh-W}, such that any
measure with the barycenter $\rho_{0}$ has no atoms in the set of
pure product states, then
$\overline{\mathrm{co}}H_{\Phi_{0}}(\rho_{0})=0$ and by the
construction of the state $\rho_{0}$ any optimal measure
$\mu_{\rho_{0}}^{\Phi_{0}}$ has no atoms.$\square$
\end{remark}

It is also shown in \cite{Sh-3} that
\begin{equation}\label{H-fun-rep-s}
\overline{\mathrm{co}}H_{\Phi}(\rho)=\mathrm{co}H_{\Phi}(\rho)=
\inf_{\{\pi_{i},
\rho_{i}\}^{\mathrm{f}}\in\mathcal{P}_{\{\rho\}}}\sum_{i}\pi_{i}H_{\Phi}(\rho_{i})
\end{equation}
in the case $H_{\Phi}(\rho)<+\infty$. By decomposing each state of
finite ensemble $\{\pi_{i}, \rho_{i}\}^{\mathrm{f}}$ into convex
countable combination of pure states and by using concavity of the
(output) entropy it is easy to obtain from (\ref{H-fun-rep-g}) and
(\ref{H-fun-rep-s}) that in the case $H_{\Phi}(\rho)<+\infty$ the
following representation holds
\begin{equation}\label{H-fun-rep-s+}
\overline{\mathrm{co}}H_{\Phi}(\rho)=\inf_{\{\pi_{i},
\rho_{i}\}\in\widehat{\mathcal{P}}_{\{\rho\}}}\sum_{i}\pi_{i}H_{\Phi}(\rho_{i}).
\end{equation}

\begin{remark}\label{hf-r-2}
Validity of representations (\ref{H-fun-rep-s}) and
(\ref{H-fun-rep-s+}) for all states with finite output entropy
implies that the definition of the entanglement of formation as
the \textup{CCoOE} of a partial trace, proposed in \cite{Sh-3}, is
\textit{reasonable} from the physical point of view. Indeed, since
in a physical experiment we can prepare only states with finite
mean energy, it is reasonable to consider in the definition of the
entanglement of formation $E_{F}(\omega)$ of a given state
$\omega$ with finite energy only convex decompositions of this
state consisting of states with finite energy. Representations
(\ref{H-fun-rep-s}) and (\ref{H-fun-rep-s+}) imply that for any
such state $\omega$ the infimum in definition (\ref{H-fun-rep-g})
of the value $\overline{\mathrm{co}}H_{\Phi}(\omega)$, where
$\Phi$ is a partial trace, can be taken only over the set of
atomic measures whose atoms are states with finite
energy.\footnote{The author is grateful to J.I.Cirac for pointing
the importance of this observation.}$\square$
\end{remark}

In the case $H_{\Phi}(\rho)=+\infty$ the general properties of the
entropy \cite{L},\cite{W} implies
$\mathrm{co}H_{\Phi}(\rho)=+\infty$, but
$\overline{\mathrm{co}}H_{\Phi}(\rho)$ may be finite (for example,
if $\Phi$ is the noiseless channel then
$\overline{\mathrm{co}}H_{\Phi}(\rho)=0$ for all states $\rho$)
and hence representation (\ref{H-fun-rep-s}) does not hold in this
case. Nevertheless this does not imply that representation
(\ref{H-fun-rep-s+}) is not true. Moreover, for all studied
examples of quantum channels $\Phi$ this representation holds for
all states $\rho$ and some general sufficient condition for its
validity can be formulated (proposition \ref{hf-p-2} below). Note
that $"\leq"$ in (\ref{H-fun-rep-s+}) is obviously follows from
(\ref{H-fun-rep-g}), but the proof of $"="$ remains an open
problem. Validity of representation (\ref{H-fun-rep-s+}) for all
channels $\Phi$ and states $\rho$ is a very desirable property
from the technical point of view since in proving general results
it is more convenient to deal with a countable sum instead of an
integral over an arbitrary probability measure. In \cite{Sh-6}
(remark 2 and the note below) it is shown that representation
(\ref{H-fun-rep-s+}) can not be proved by using only such
analytical properties of the output entropy as concavity and lower
semicontinuity.

For brevity and according to the tradition the convex closure
$\overline{\mathrm{co}}H_{\Phi}$ of the output entropy $H_{\Phi}$
of a channel $\Phi$ will be denoted below by $\hat{H}_{\Phi}$.

The $\chi$-function of the channel $\Phi$ is defined by
(cf.\cite{H-Sh-1},\cite{H-Sh-2})
\begin{equation}\label{chi-fun-def}
\chi_{\Phi}(\rho)=\sup_{\{\pi_{i},
\rho_{i}\}^{\mathrm{f}}\in\mathcal{P}_{\{\rho\}}}
\sum_{i}\pi_{i}H(\Phi(\rho _{i})\|\Phi(\rho)).
\end{equation}

The $\chi$-function of an arbitrary infinite dimensional channel
$\Phi$ is a nonnegative concave and lower semicontinuous function,
such that
\begin{equation}\label{chi-exp-2}
\chi _{\Phi}(\rho)=H_{\Phi}(\rho)-\hat{H}_{\Phi}(\rho)
\end{equation}
for all states $\rho$ with finite output entropy $H_{\Phi}(\rho)$
\cite{Sh-2}.

We will denote by $h_{2}(p)$ the binary entropy $-p\log
p-(1-p)\log(1-p)$.

\section{Continuity properties}

For arbitrary channel $\Phi$ the function
$\hat{H}_{\Phi}=\overline{\mathrm{co}}H_{\Phi}$ is lower
semicontinuous by definition but it is not continuous in general.
Nevertheless the function $\hat{H}_{\Phi}$ can have continuous
restrictions to some sets of states. Moreover for some channels
$\Phi$ it can be continuous and bounded on the whole state space
(for example, if $\Phi$ is the noiseless channel then
$\hat{H}_{\Phi}\equiv 0$). In this section we consider two
continuity conditions for this function.

In \cite{Sh-3} the sufficient condition of continuity of the
restriction of the function $\hat{H}_{\Phi}$ to a subset of
$\mathfrak{S}(\mathcal{H})$ is obtained (proposition 7). This
condition can be reformulated as follows.

\begin{property}\label{hf-p-1}
\textit{Let
$\Phi:\mathfrak{S}(\mathcal{H})\mapsto\mathfrak{S}(\mathcal{H}')$
be an arbitrary quantum channel. If  $\{\rho_{n}\}$ is a sequence
of states converging to the state $\rho_{0}$ such that
$\lim_{n\rightarrow+\infty}H_{\Phi}(\rho_{n})=H_{\Phi}(\rho_{0})<+\infty$
then
$\lim_{n\rightarrow+\infty}\hat{H}_{\Phi}(\rho_{n})=\hat{H}_{\Phi}(\rho_{0})$.}
\end{property}

\begin{remark}\label{hf-r-7} The statement of proposition \ref{hf-p-1} seems
surprising by the following reason. A value of the output entropy
$H_{\Phi}$ at a particular state $\rho$ is completely defined by
the output state $\Phi(\rho)$ and it does not depend on the action
of the channel $\Phi$ on other input states. Hence the condition
$\lim_{n\rightarrow+\infty}H_{\Phi}(\rho_{n})=H_{\Phi}(\rho_{0})$
depends only on the action of the channel $\Phi$ on the states of
the sequence $\{\rho_{n}\}$ and its limit state $\rho_{0}$ so that
it is a \textit{local} condition. In contrast to this a value of
the function $\hat{H}_{\Phi}$ at a particular state $\rho$ depends
on the action of the channel $\Phi$ on the whole state
space.\footnote{More precisely, on the union of supports of all
measures with the barycenter $\rho$.} It follows as from the
general definition of the convex closure as from representation
(\ref{H-fun-rep-g}). Hence the property
$\lim_{n\rightarrow+\infty}\hat{H}_{\Phi}(\rho_{n})=\hat{H}_{\Phi}(\rho_{0})$
depends on the action of the channel $\Phi$ on the whole state
space so that it is \textit{nonlocal}. Nevertheless proposition
\ref{hf-p-1} provides a sufficient condition of its validitity in
terms of the above local condition. $\square$
\end{remark}

It is shown in \cite{Sh-3} that proposition \ref{hf-p-1} implies
continuity of the entanglement of formation on the set of all states
of bipartite system with bounded mean energy. In this paper we will
use this proposition in proving our basic approximation result
(lemma \ref{hf-l-1} in section 4), concerning the additivity
problem.

Note also that proposition \ref{hf-p-1} implies the following
observation, which shows continuity of the \textup{CCoOE} with
respect to the convergence defined by the quantum relative entropy
$H(\cdot\|\cdot)$ \cite{L}.

\begin{corollary}\label{hf-c-7}
\textit{Let
$\Phi:\mathfrak{S}(\mathcal{H})\mapsto\mathfrak{S}(\mathcal{H}')$
be a quantum channel and $\rho_{0}$ be a state in
$\mathfrak{S}(\mathcal{H})$ such that
$\mathrm{Tr}(\Phi(\rho_{0}))^{\lambda}<+\infty$ for some
$\lambda<1$. Then
$\lim_{n\rightarrow+\infty}\hat{H}_{\Phi}(\rho_{n})=\hat{H}_{\Phi}(\rho_{0})$
for arbitrary sequence $\{\rho_{n}\}$ of states in
$\mathfrak{S}(\mathcal{H})$ such that
$\lim_{n\rightarrow+\infty}H(\rho_{n}\|\rho_{0})=0$.}
\end{corollary}

\textbf{Proof.} By proposition \ref{hf-p-1} it is sufficient to
show that
$\lim_{n\rightarrow+\infty}H_{\Phi}(\rho_{n})=H_{\Phi}(\rho_{0})<+\infty$.
But this follows from proposition 2 in \cite{Sh-4} since by
monotonicity of the relative entropy the condition
$\lim_{n\rightarrow+\infty}H(\rho_{n}\|\rho_{0})=0$ implies
$\lim_{n\rightarrow+\infty}H(\Phi(\rho_{n})\|\Phi(\rho_{0}))=0$.
$\square$

Note that the conditions of corollary \ref{hf-c-7} are valid for
arbitrary Gaussian channel $\Phi$ and arbitrary Gaussian state
$\rho_{0}$.

The assertion of proposition \ref{hf-p-1} can not be converted as
it follows from the example of the noiseless channel $\Phi$.

In contrast to proposition \ref{hf-p-1} the following proposition
provides a necessary and sufficient condition of continuity of the
\textup{CCoOE} on the whole state space.

\begin{property}\label{hf-p-2}
\textit{Let
$\Phi:\mathfrak{S}(\mathcal{H})\mapsto\mathfrak{S}(\mathcal{H}')$
be an arbitrary quantum channel.}

\textit{The convex closure $\hat{H}_{\Phi}$ of the output entropy
$H_{\Phi}$ is bounded and continuous on the set
$\mathfrak{S}(\mathcal{H})$ if and only if the output entropy
$H_{\Phi}$ is bounded and continuous on the set
$\mathrm{ext}\mathfrak{S}(\mathcal{H})$. In this case
representation (\ref{H-fun-rep-s+}) holds for arbitrary state
$\rho$ in $\mathfrak{S}(\mathcal{H})$.}

\textit{The above equivalent properties hold if there exist
separable Hilbert space $\mathcal{K}$ and set
$\{\alpha_{\psi}\}_{\psi\in\mathcal{H},\|\psi\|=1}$ of isomorphism
from $\mathfrak{S}(\mathcal{H}')$ onto $\mathfrak{S}(\mathcal{K})$
such that $\alpha_{\psi}(\Phi(|\psi\rangle\langle\psi|))$ lies for
each $\psi$ in some compact subset $\mathcal{A}$ of
$\mathfrak{S}(\mathcal{K})$, on which the entropy is continuous.}
\end{property}

\textbf{Proof} The first part of this proposition follows from
proposition 5 and corollary 9 in \cite{Sh-6}. The second one
follows from corollary \ref{hf-cont-cond-1} in the Appendix 5.1,
where general condition of continuity of the entropy is
considered.$\square$

Note that continuity of the output entropy on the set of pure
states is not very restrictive requirement. Indeed, this
continuity trivially holds for noiseless channel, for which the
output entropy coincides with the entropy of a state and is far
from being continuous on the whole state space. Note that in this
case the sufficient condition in proposition \ref{hf-p-2} is also
trivially verified with $\mathcal{K}=\mathcal{H}'$,
$\mathcal{A}=\{|\psi_{0}\rangle\langle\psi_{0}|\}$ and
$\alpha_{\psi}=U_{\psi}(\cdot)U_{\psi}^{*}$, where $U_{\psi}$ is
any unitary such that $U_{\psi}|\psi\rangle=|\psi_{0}\rangle$ and
$|\psi_{0}\rangle$ is some fixed unit vector in $\mathcal{H}'$.

The nontrivial application of the sufficient condition in
proposition \ref{hf-p-2} is the proof of continuity of the
\textup{CCoOE} for the class of channels considered in the following
example.

\textbf{Example.} Let $\mathcal{H}_{a}$ be the Hilbert space
$\mathcal{L}_{2}([-a,+a])$, where $a\leq+\infty$ and
$\{U_{t}\}_{t\in\mathbb{R}}$ be the group of unitary operators in
$\mathcal{H}_{a}$ defined by
$$
(U_{t}\varphi)(x)=\exp(-\mathrm{i}tx)\varphi(x),\quad\forall\varphi\in\mathcal{H}_{a}.
$$
For given probability density function $p(t)$ consider the channel
$$
\Phi_{p}^{a}:\mathfrak{S}(\mathcal{H}_{a})\ni\rho\mapsto\int_{-\infty}^{+\infty}U_{t}\rho
U_{t}^{*}p(t)dt\in\mathfrak{S}(\mathcal{H}_{a}).
$$

Assume that the differential entropy
$\int_{-\infty}^{+\infty}p(t)(-\log p(t))dt$ of the distribution
$p(t)$ is finite and that the function $p(t)$ is bounded and
monotonous on $(-\infty,-b]$ and on $[+b,+\infty)$ for
sufficiently large $b$\footnote{The last assumption is for
technical convenience.}. By using the sufficient condition in
proposition \ref{hf-p-2} it is possible to show (see Appendix 5.2
for details) that:
\begin{itemize}
  \item if $a<+\infty$ then the equivalent properties in proposition \ref{hf-p-2} hold for the channel $\Phi_{p}^{a}$;
  \item if $a=+\infty$ then for the channel $\Phi_{p}^{a}$ there exists pure state in
  $\mathfrak{S}(\mathcal{H}_{a})$ with infinite output entropy.
 \end{itemize}

Since for each $a>0$ the space $\mathcal{H}_{a}$ can be considered
as a subspace of $\mathcal{H}_{+\infty}$ the set
$\mathfrak{S}(\mathcal{H}_{a})$ can be considered as a subset of
$\mathfrak{S}(\mathcal{H}_{+\infty})$. The restriction of the
channel $\Phi_{p}^{+\infty}$ to the subset
$\mathfrak{S}(\mathcal{H}_{a})$ coincides with the channel
$\Phi_{p}^{a}$ and hence the restriction of the function
$\hat{H}_{\Phi_{p}^{+\infty}}$ to the subset
$\mathfrak{S}(\mathcal{H}_{a})$ coincides with the function
$\hat{H}_{\Phi_{p}^{a}}$. By the above observation the unbounded
function $\hat{H}_{\Phi_{p}^{+\infty}}$ has bounded and continuous
restriction to each subset from the increasing family
$\{\mathfrak{S}(\mathcal{H}_{a})\}_{a>0}$ such that
$\overline{\bigcup_{a>0}\mathfrak{S}(\mathcal{H}_{a})}=\mathfrak{S}(\mathcal{H}_{+\infty})$.

\section{The additivity problem}

Let $\Phi :\mathfrak{S}(\mathcal{H})\mapsto
\mathfrak{S}(\mathcal{H}^{\prime })$ and
$\Psi:\mathfrak{S}(\mathcal{K})\mapsto
\mathfrak{S}(\mathcal{K}^{\prime})$ be two channels. In this section
we consider the superadditivity property of the \textup{CCoOE},
which means validity of the inequality
\begin{equation}\label{super-add}
\hat{H}_{\Phi\otimes\Psi}(\omega )\geq \hat{H}_{\Phi}
(\omega^{\mathcal{H}} )+\hat{H}_{\Psi} (\omega^{\mathcal{K}})
\end{equation}
for all states $\omega$ in
$\mathfrak{S}(\mathcal{H}\otimes\mathcal{K})$, where
$\omega^{\mathcal{H}}=\mathrm{Tr}_{\mathcal{K}}\omega$ and
$\omega^{\mathcal{K}}=\mathrm{Tr}_{\mathcal{H}}\omega$. If $\Phi$
and $\Psi$ are partial trace channels then this property means the
strong superadditivity of the entanglement of formation
\cite{A&B}. Hence, as it is proved by P.Shor \cite{Shor-e-a-q},
the global additivity conjecture in the finite dimensional case
can be expressed as the superadditivity of the \textup{CCoOE} for
partial trace channels.

The superadditivity of the \textup{CCoOE} for given two finite
dimensional channels is equivalent to the additivity of the Holevo
capacity for these channels with arbitrary constraints, which can
be expressed as the subadditivity property of the $\chi$-function,
consisting in validity of the inequality
\begin{equation}\label{sub-add}
\chi _{\Phi \otimes \Psi }(\omega)\leq
\chi_{\Phi}(\omega^{\mathcal{H}})+\chi_{\Psi}(\omega^{\mathcal{K}})
\end{equation}
for all states $\omega$ in
$\mathfrak{S}(\mathcal{H}\otimes\mathcal{K})$ \cite{H-Sh-1}.

The continuity condition for the \textup{CCoOE}, considered in
section 3, implies the following approximation result, which plays
a basic role in this section.

\begin{lemma}\label{hf-l-1}
\textit{Let
$\Phi:\mathfrak{S}(\mathcal{H})\mapsto\mathfrak{S}(\mathcal{H}')$
and
$\Psi:\mathfrak{S}(\mathcal{K})\mapsto\mathfrak{S}(\mathcal{K}')$
be quantum channels. If for arbitrary finite rank state
$\omega_{0}$ in $\mathfrak{S}(\mathcal{H}\otimes\mathcal{K})$ such
that $H_{\Phi\otimes\Psi}(\omega_{0})<+\infty$ there exists
sequence $\{\omega_{n}\}$ of states in
$\mathfrak{S}(\mathcal{H}\otimes\mathcal{K})$ such that}
\begin{itemize}
  \item \textit{$\lim\limits_{n\rightarrow+\infty}\omega_{n}=\omega_{0}\;$ and
  $\;\lim\limits_{n\rightarrow+\infty}H_{\Phi\otimes\Psi}(\omega_{n})=H_{\Phi\otimes\Psi}(\omega_{0})$,}
  \item \textit{inequality (\ref{super-add}) holds with $\omega=\omega_{n}$ for each $n\in\mathbb{N}$,}
\end{itemize}
\textit{then inequality (\ref{super-add}) holds for all states
$\omega$ in $\mathfrak{S}(\mathcal{H}\otimes\mathcal{K})$.}
\end{lemma}

\textbf{Proof.} Note first that the condition of the lemma implies
inequality (\ref{super-add}) for arbitrary finite rank state
$\omega$ such that $H_{\Phi\otimes\Psi}(\omega)<+\infty$. Let
$\omega_{0}$ be such a state. By proposition \ref{hf-p-1} the
first property of the corresponding sequence $\{\omega_{n}\}$
implies
$\lim_{n\rightarrow+\infty}\hat{H}_{\Phi\otimes\Psi}(\omega_{n})=\hat{H}_{\Phi\otimes\Psi}(\omega_{0})$.
By this and due to lower semicontinuity of the functions
$\hat{H}_{\Phi}$ and $\hat{H}_{\Psi}$ the second property of the
sequence $\{\omega_{n}\}$ implies inequality (\ref{super-add})
with $\omega=\omega_{0}$.

It is sufficient to prove inequality (\ref{super-add}) for an
arbitrary state $\omega$ in
$\mathfrak{S}(\mathcal{H}\otimes\mathcal{K})$ such that
$\hat{H}_{\Phi\otimes\Psi}(\omega)<+\infty$. Let $\omega_{0}$ be
such a state. By lemma 3 in \cite{Sh-3} there exists a sequence
$\{\omega_{n}\}$ of finite rank states in
$\mathfrak{S}(\mathcal{H}\otimes\mathcal{K})$ converging to the
state $\omega_{0}$ such that
\begin{equation}\label{f-1}
H_{\Phi\otimes\Psi}(\omega_{n})<+\infty,\;\forall n\quad
\textup{and}\quad
\lim_{n\rightarrow+\infty}\hat{H}_{\Phi\otimes\Psi}(\omega_{n})=\hat{H}_{\Phi\otimes\Psi}(\omega_{0}).
\end{equation}

By the previous observation inequality (\ref{super-add}) holds
with $\omega=\omega_{n}$ for all $n=1,2,..$. This, lower
semicontinuity of the functions $\hat{H}_{\Phi}$ and
$\hat{H}_{\Psi}$ and (\ref{f-1}) imply inequality
(\ref{super-add}) with $\omega=\omega_{0}$.$\square$
\begin{remark}\label{hf-r-2}
By lemma \ref{hf-l-1} to prove the superadditivity of the
\textup{CCoOE} for given two channels $\Phi$ and $\Psi$ it is
sufficient to prove inequality (\ref{super-add}) for arbitrary
finite rank state $\omega$ in
$\mathfrak{S}(\mathcal{H}\otimes\mathcal{K})$ such that
$H_{\Phi\otimes\Psi}(\omega)<+\infty$.$\square$
\end{remark}

In \cite{Sh-2} it is proved that additivity of the Holevo capacity
for all finite dimensional channels implies additivity of the
Holevo capacity for all infinite dimensional channels with
arbitrary constraints. Lemma \ref{hf-l-1} makes possible to prove
the analogous result, concerning the superadditivity of the
\textup{CCoOE}.

\begin{theorem}\label{hf-th-1}
\textit{If the superadditivity of the \textup{CCoOE} holds for all
finite dimensional channels then the superadditivity of the
\textup{CCoOE} holds for all infinite dimensional channels.}
\end{theorem}

\textbf{Proof.} By the observation in \cite{MSW} any channel is
unitarily equivalent to a particular
subchannel\footnote{Subchannel of a channel is a restriction of
this channel to the set of all states supported by a particular
subspace.} of a partial trace. Since the superadditivity of the
\textup{CCoOE} for arbitrary two channels implies the same
property for any their subchannels it is sufficient to consider
the case of partial trace channels $\Phi$ and $\Psi$. Let
$\mathcal{H},\mathcal{L},\mathcal{K},\mathcal{N}$ be separable
Hilbert spaces and
$$
\Phi(\rho)=\mathrm{Tr}_{\mathcal{L}}\rho,\;\rho\in\mathfrak{S}(\mathcal{H}\otimes\mathcal{L})\quad\textup{and}\quad
\Psi(\sigma)=\mathrm{Tr}_{\mathcal{N}}\sigma,\;\sigma\in\mathfrak{S}(\mathcal{K}\otimes\mathcal{N}).
$$
Let $\omega_{0}$ be a state in
$\mathfrak{S}(\mathcal{H}\otimes\mathcal{L}\otimes\mathcal{K}\otimes\mathcal{N})$
such that $H_{\Phi\otimes\Psi}(\omega_{0})<+\infty$. Let
$\{P_{n}\},\{R_{n}\},\{Q_{n}\},\{S_{n}\}$ be increasing sequences of
$n$-dimensional projectors in the spaces
$\mathcal{H},\mathcal{L},\mathcal{K},\mathcal{N}$, strongly
converging to the identity operators
$I_{\mathcal{H}},I_{\mathcal{L}},I_{\mathcal{K}},I_{\mathcal{N}}$
correspondingly. Consider the sequence of states
$$
\omega_{n}=\frac{W_{n}\omega_{0}W_{n}}{\mathrm{Tr}W_{n}\omega_{0}},\quad
\textup{where}\quad W_{n}=P_{n}\otimes R_{n}\otimes Q_{n}\otimes
S_{n},
$$
converging to the state $\omega_{0}$.

For each $n$ the following operator inequality holds
\begin{equation}\label{oper-ineq}
\begin{array}{c}
\Phi\otimes\Psi(\omega_{n})=\mathrm{Tr}_{\mathcal{L}\otimes\mathcal{N}}\omega_{n}\\\\\leq
(\mathrm{Tr}W_{n}\omega_{0})^{-1}P_{n}\otimes
Q_{n}\cdot(\mathrm{Tr}_{\mathcal{L}\otimes\mathcal{N}}\omega_{0})\cdot
P_{n}\otimes Q_{n}\\\\=(\mathrm{Tr}W_{n}\omega_{0})^{-1}P_{n}\otimes
Q_{n}\cdot \Phi\otimes\Psi(\omega_{0})\cdot P_{n}\otimes Q_{n}.
\end{array}
\end{equation}

Indeed, let $\{|i\rangle\}$ and $\{|j\rangle\}$ be ONB in the spaces
$\mathcal{L}$ and $\mathcal{N}$ correspondingly such that
$$
R_{n}=\sum_{i=1}^{n}|i\rangle\langle i|\quad \textup{and} \quad
S_{n}=\sum_{j=1}^{n}|j\rangle\langle j|.
$$
For arbitrary $|\varphi\rangle$ in $P_{n}\otimes
Q_{n}(\mathcal{H}\otimes\mathcal{K})$ we have
$$
\begin{array}{c}
\displaystyle\langle\varphi|\mathrm{Tr}_{\mathcal{L}\otimes\mathcal{N}}\omega_{n}|\varphi\rangle=
\sum_{i,j=1}^{+\infty}\langle\varphi\otimes i \otimes
j\,|\,\omega_{n}|\varphi\otimes i \otimes j\rangle=\\\\
\displaystyle(\mathrm{Tr}W_{n}\omega_{0})^{-1}\sum_{i,j=1}^{n}\langle\varphi\otimes
i \otimes j\,|\,\omega_{0}|\varphi\otimes i \otimes
j\rangle\leq(\mathrm{Tr}W_{n}\omega_{0})^{-1}
\langle\varphi|\mathrm{Tr}_{\mathcal{L}\otimes\mathcal{N}}\omega_{0}|\varphi\rangle,
\end{array}
$$
which implies (\ref{oper-ineq}).

Since inequality (\ref{oper-ineq}) means decomposition
$$
P_{n}\otimes Q_{n}\cdot \Phi\otimes\Psi(\omega_{0})\cdot
P_{n}\otimes
Q_{n}=\lambda\Phi\otimes\Psi(\omega_{n})+(1-\lambda)(\textup{positive
operator}),
$$
where $\lambda=\mathrm{Tr}W_{n}\omega_{0}$, concavity of the entropy
and proposition 4 in \cite{L} imply
$$
\begin{array}{c}
H(\Phi\otimes\Psi(\omega_{n}))\leq(\mathrm{Tr}W_{n}\omega_{0})^{-1}H(P_{n}\otimes
Q_{n}\cdot \Phi\otimes\Psi(\omega_{0})\cdot P_{n}\otimes
Q_{n})\\\\\leq(\mathrm{Tr}W_{n}\omega_{0})^{-1}H(\Phi\otimes\Psi(\omega_{0}))<+\infty.
\end{array}
$$
It follows from this and lower semicontinuity of the entropy that
$$
\lim_{n\rightarrow+\infty}H(\Phi\otimes\Psi(\omega_{n}))=H(\Phi\otimes\Psi(\omega_{0}))<+\infty.
$$

By the assumption inequality (\ref{super-add}) holds with
$\omega=\omega_{n}$ for each $n$.

Lemma \ref{hf-l-1} implies the superadditivity of the \textup{CCoOE}
for the channels $\Phi$ and $\Psi$. $\square$

By using corollary 3 in \cite{F} theorem \ref{hf-th-1} can be
strengthened as follows.
\begin{corollary}\label{hf-c-10}
\textit{If the superadditivity of the \textup{CCoOE} holds for all
finite dimensional unital channels then the superadditivity of the
\textup{CCoOE} holds for all infinite dimensional channels.}
\end{corollary}

By the observation in \cite{MSW} theorem \ref{hf-th-1} can be
reformulated in terms of the strong superadditivity of the
entanglement of formation.

\begin{corollary}\label{hf-c-11}
\textit{The strong superadditivity of the \textup{EoF} in the finite
dimensional case implies the strong superadditivity of the
\textup{EoF} in the infinite dimensional case.}\footnote{In the last
case we use the definition of the \textup{EoF} as the
\textup{\textup{CCoOE}} of a partial trace \cite{Sh-3}.}
\end{corollary}

Since the superadditivity of the \textup{CCoOE} for two channels
obviously implies the additivity of the minimal output entropy for
these channels theorem \ref{hf-th-1}, Shor's theorem
\cite{Shor-e-a-q} and corollary 3 in \cite{F} imply the following
result.

\begin{corollary}\label{hf-c-12}
\textit{The additivity of the minimal output entropy for all finite
dimensional unital channels implies the additivity of the minimal
output entropy for all infinite dimensional channels.}
\end{corollary}

Theorem \ref{hf-th-1} and theorem 3 in \cite{Sh-2} provide
infinite dimensional generalization of Shor's theorem
\cite{Shor-e-a-q}.

\begin{theorem}\label{hf-th-2}
\textit{The following properties are equivalent:}
\begin{itemize}
  \item \textit{the additivity of the Holevo capacity holds for all infinite dimensional
  channels with arbitrary constraints;}
  \item \textit{the additivity of the minimal output entropy holds for all infinite dimensional
  channels;}
  \item \textit{the strong superadditivity of the \textup{EoF} holds for arbitrary state in infinite dimensional
bipartite system.}
\end{itemize}
\end{theorem}

Lemma \ref{hf-l-1}, proposition 7 and theorem 2 in \cite{Sh-2}
make possible to derive the superadditivity of the \textup{CCoOE}
for nontrivial classes of infinite dimensional channels from the
corresponding finite dimensional results
\cite{Shor-e-b-c},\cite{H-Sh-1}.

\begin{property}\label{hf-p-3}
\textit{Let $\Psi$ be an arbitrary channel. The superadditivity
the \textup{CCoOE} holds in each of the following cases: }

$\mathrm{(i)}$ $\Phi $\textit{\ is the noiseless channel;}

$\mathrm{(ii)}$ $\Phi $\textit{\ is an entanglement breaking
channel;}

$\mathrm{(iii)}$ $\Phi $\textit{\ is a direct sum mixture
(cf.\cite{H-Sh-1}) of a noiseless channel and a channel $\Phi_{0}$
such that the superadditivity the \textup{CCoOE} holds for
$\Phi_{0}$ and $\Psi$ (in particular, an entanglement breaking
channel).}
\end{property}

\textbf{Proof.} $\mathrm{(i)}$ \footnote{This part of the
proposition can be also proved by using the notion of complementary
channel described below.} Let $\Phi=\mathrm{Id}$ be the noiseless
channel. Let $\omega_{0}$ be a state such that
$H_{\Phi\otimes\Psi}(\omega_{0})<+\infty$  and $\{P_{n}\}$ be the
increasing sequence of spectral projectors of the state
$\omega_{0}^{\mathcal{H}}=\mathrm{Tr}_{\mathcal{K}}\omega_{0}$.
Consider the sequence of states
$$
\omega_{n}=\lambda_{n}^{-1}P_{n}\otimes
I\,\omega_{0}\,P_{n}\otimes I,
$$
converging to the state $\omega_{0}$, where
$\lambda_{n}=\mathrm{Tr}(P_{n}\otimes I\,\omega_{0})$.

Since $\Phi\otimes\Psi(\omega_{n})=\lambda_{n}^{-1}P_{n}\otimes
I\,\Phi\otimes\Psi(\omega_{0})\,P_{n}\otimes I$ for all $n$ these
states have finite entropy and Simon's convergence theorem for
entropy \cite{Simon} implies
$$
\lim_{n\rightarrow+\infty}H_{\Phi\otimes\Psi}(\omega_{n})=H_{\Phi\otimes\Psi}(\omega_{0})<+\infty.
$$

Since $\omega_{n}^{\mathcal{H}}=\Phi(\omega_{n}^{\mathcal{H}})$ is a
finite rank state $H_{\Phi}(\omega_{n}^{\mathcal{H}})<+\infty$. This
and well known inequality
$$
|H_{\Phi}(\omega_{n}^{\mathcal{H}})-H_{\Psi}(\omega_{n}^{\mathcal{K}})|\leq
H_{\Phi\otimes\Psi}(\omega_{n})< +\infty
$$
implies $H_{\Psi}(\omega_{n}^{\mathcal{K}})<+\infty$. By
proposition 7 and theorem 2 in \cite{Sh-2} inequality
(\ref{super-add}) holds with $\omega=\omega_{n}$ for each $n$.

Thus the sequence $\{\omega_{n}\}$ satisfies the both conditions
in lemma \ref{hf-l-1} for given state $\omega_{0}$. By this lemma
inequality (\ref{super-add}) holds for all states $\omega$ in
$\mathfrak{S}(\mathcal{H}\otimes\mathcal{K})$.

$\mathrm{(ii)}$ Let $\Phi$ be an entanglement-breaking channel. By
lemma \ref{hf-l-1} it is sufficient to prove inequality
(\ref{super-add}) for all states $\omega$ such that
$H_{\Phi\otimes\Psi}(\omega)<+\infty$. Let $\omega_{0}$ be such a
state. Since the channel $\Phi$ is entanglement-breaking the state
$\Phi\otimes\Psi(\omega_{0})$ is separable and hence
$$
\max\{H_{\Phi}(\omega_{0}^{\mathcal{H}}),H_{\Psi}(\omega_{0}^{\mathcal{K}})\}\leq
H_{\Phi\otimes\Psi}(\omega_{0})<+\infty.
$$
By proposition 7 and theorem 2 in \cite{Sh-2} inequality
(\ref{super-add}) holds with $\omega=\omega_{0}$.

$\mathrm{(iii)}$ Let $\Phi_{q}=q\mathrm{Id}\oplus (1-q)\Phi_{0}$.
Note that
$H_{\Phi_{q}}=qH_{\mathrm{Id}}+(1-q)H_{\Phi_{0}}+h_{2}(q)$ and
$H_{\Phi_{q}\otimes\Psi}=qH_{\mathrm{Id}\otimes\Psi}+(1-q)H_{\Phi_{0}\otimes\Psi}+h_{2}(q)$.
By using this, $\mathrm{(i)}$ and the condition on the channel
$\Phi_{0}$ we obtain
$$
\begin{array}{c}
 \hat{H}_{\Phi_{q}\otimes\Psi}(\omega)\geq
q\hat{H}_{\mathrm{Id}\otimes\Psi}(\omega)+(1-q)\hat{H}_{\Phi_{0}\otimes\Psi}(\omega)+h_{2}(q)\\\\\geq
q\hat{H}_{\Psi}(\omega^{\mathcal{K}})+(1-q)(\hat{H}_{\Phi_{0}}(\omega^{\mathcal{H}})+
\hat{H}_{\Psi}
(\omega^{\mathcal{K}}))+h_{2}(q)=\hat{H}_{\Phi_{q}}(\omega^{\mathcal{H}})+
\hat{H}_{\Psi} (\omega^{\mathcal{K}}),
\end{array}
$$
for any state $\omega$ in
$\mathfrak{S}(\mathcal{H}\otimes\mathcal{K})$, where the last
equality follows from the observation that the infimum in
expression (\ref{H-fun-rep-g}) for the value
$\hat{H}_{\mathrm{Id}}(\omega^{\mathcal{H}})=0$ is achieved at any
measure supported by pure states and having the barycenter
$\omega^{\mathcal{H}}$ . $\square$

Proposition \ref{hf-p-3} implies in particular the following result,
which seems to be nontrivial in the infinite dimensional case.

\begin{corollary}\label{hf-c-1}
\textit{For arbitrary channels $\Phi$ and $\Psi$ the following
inequality holds
$$
\hat{H}_{\Phi\otimes\Psi}(\omega)\geq
\max\left\{\hat{H}_{\Phi}(\omega^{\mathcal{H}}), \hat{H}_{\Psi}
(\omega^{\mathcal{K}})\right\},\quad\forall
\omega\in\mathfrak{S}(\mathcal{H}\otimes\mathcal{K}),
$$
which implies in particular that
$$
H_{\mathrm{min}}(\Phi\otimes\Psi)\geq
\max\left\{H_{\mathrm{min}}(\Phi), H_{\mathrm{min}}(\Psi)\right\}.
$$}
\end{corollary}

\textbf{Proof.} By proposition \ref{hf-p-3}
$\hat{H}_{\Phi\otimes\mathrm{Id}}(\omega)\geq\hat{H}_{\Phi}(\omega^{\mathcal{H}})$
for any state $\omega$. By using this and noting that
$\hat{H}_{\Phi\otimes\mathrm{Id}}(\mathrm{Id}\otimes\Psi(\omega))\leq\hat{H}_{\Phi\otimes\Psi}(\omega)$
for any state $\omega$ (this follows from expression
(\ref{H-fun-rep-g}) for the \textup{CCoOE}) we obtain the first
assertion of the corollary. The second one obviously follows from
the first. $\square$

\begin{remark}\label{hf-r-3}
By corollary \ref{hf-c-1} if the output entropy of one of the
channels $\Phi$ and $\Psi$ is everywhere infinite then the output
entropy of the channel $\Phi\otimes\Psi$ is everywhere infinite as
well despite the properties of the another channel. $\square$
\end{remark}

In \cite{H-comp-ch},\cite{R-comp-ch} it is noted that the notion
of complementary channel introduced in \cite{D&Sh} is very useful
in connection with the additivity problem.

If
$\Phi:\mathfrak{S}(\mathcal{H})\mapsto\mathfrak{S}(\mathcal{H}')$
is an arbitrary channel then by the Stinespring dilation theorem
there exist a Hilbert space $\mathcal{H}''$ and an isometry
$V:\mathcal{H}\mapsto\mathcal{H}'\otimes\mathcal{H}''$ such that
$$
\Phi(\rho)=\mathrm{Tr}_{\mathcal{H}''}V\rho V^{*},\quad
\forall\rho\in\mathfrak{S}(\mathcal{H}).
$$
The channel
$\widehat{\Phi}:\mathfrak{S}(\mathcal{H})\mapsto\mathfrak{S}(\mathcal{H}'')$
defined by
$$
\widehat{\Phi}(\rho)=\mathrm{Tr}_{\mathcal{H}'}V\rho V^{*},\quad
\forall\rho\in\mathfrak{S}(\mathcal{H}),
$$
is called complementary to the channel $\Phi$. The complementary
channel is defined uniquely up to unitary equivalence.

The basic properties of the complementary relation are the
following:
\begin{itemize}
  \item
  $H_{\Phi}(\rho)=H_{\widehat{\Phi}}(\rho)$ for arbitrary channel
  $\Phi$ and any pure
  state $\rho$;
  \item
  $\widehat{\Phi\otimes\Psi}=\widehat{\Phi}\otimes\widehat{\Psi}$ for arbitrary channels
  $\Phi$ and $\Psi$.
\end{itemize}

The first property and the definition of the \textup{CCoOE} imply
$\hat{H}_{\Phi}\equiv\hat{H}_{\widehat{\Phi}}$ despite the fact
that  $H_{\Phi}$ and $H_{\widehat{\Phi}}$ may be substantially
different functions. This and the second property show that the
superadditivity of the \textup{CCoOE} for channels $\Phi$ and
$\Psi$ is equivalent to the same property for the complementary
channels $\widehat{\Phi}$ and $\widehat{\Psi}$ (theorem 1 in
\cite{H-comp-ch}).

In the finite dimensional case the superadditivity of the
\textup{CCoOE} for particular channels $\Phi$ and $\Psi$ is
equivalent to additivity of the Holevo capacity for these channels
with arbitrary constraints. Hence in this case the above observation
can be formulated in terms of this property: additivity of the
Holevo capacity for channels $\Phi$ and $\Psi$ with arbitrary
constraints is equivalent to the same property for the complementary
channels $\widehat{\Phi}$ and $\widehat{\Psi}$. Note that in
contrast to the \textup{CCoOE} the Holevo capacities for channels
related by the complementary relation may be substantially
different, but the common \textup{CCoOE} plays, loosely speaking,
the role of bridge between the additivity properties of the Holevo
capacity for pairs of complementary channels.

In the infinite dimensional case there exist only conditional
relations between the above additivity properties (theorem 2 in
\cite{Sh-2}) and hence we can not prove the infinite dimensional
version of the above observation, concerning the additivity of the
Holevo capacity for pairs of complementary channels. Nevertheless
the following conditional version can be proved.
\begin{property}\label{hf-p-4}
\textit{Let
$\Phi:\mathfrak{S}(\mathcal{H})\mapsto\mathfrak{S}(\mathcal{H}')$
and
$\Psi:\mathfrak{S}(\mathcal{K})\mapsto\mathfrak{S}(\mathcal{K}')$
be such channels that
$$
H_{\Phi}(\rho)<+\infty,\;\forall\rho\in\mathrm{ext}\mathfrak{S}(\mathcal{H})\quad
and\quad
H_{\Psi}(\sigma)<+\infty,\;\forall\sigma\in\mathrm{ext}\mathfrak{S}(\mathcal{K}).
$$}
\textit{Then the additivity of the Holevo capacity for the channels
$\Phi$ and $\Psi$ with arbitrary constraints is equivalent to the
same property for the complementary channels $\widehat{\Phi}$ and
$\widehat{\Psi}$.}
\end{property}

\textbf{Proof.} By the first property of the complementary relation
the condition of the proposition is symmetrical with respect to the
pairs $(\Phi,\Psi)$ and $(\widehat{\Phi},\widehat{\Psi})$. Hence it
is sufficient to show that the additivity of the Holevo capacity for
channels $\Phi$ and $\Psi$ with arbitrary constraints implies the
same property for the complementary channels $\widehat{\Phi}$ and
$\widehat{\Psi}$.

Let $\mathfrak{S}_{\mathrm{f}}(\mathcal{H}\otimes\mathcal{K})$ be
the subset of $\mathfrak{S}(\mathcal{H}\otimes\mathcal{K})$
consisting of all states $\omega$ such that
$\mathrm{rank}\,\omega^{\mathcal{H}}<+\infty$ and
$\mathrm{rank}\,\omega^{\mathcal{K}}<+\infty$. Note that the
condition of the proposition implies
$H_{\Phi}(\omega^{\mathcal{H}})<+\infty$ and
$H_{\Psi}(\omega^{\mathcal{K}})<+\infty$ for arbitrary state
$\omega\in\mathfrak{S}_{\mathrm{f}}(\mathcal{H}\otimes\mathcal{K})$.
By the part $\mathrm{(i)}\Rightarrow\mathrm{(iii)}$ of theorem 2
in \cite{Sh-2} the additivity of the Holevo capacity for the
channels $\Phi$ and $\Psi$ with arbitrary constraints implies
validity of inequality (\ref{super-add}) for these channels and
arbitrary state
$\omega\in\mathfrak{S}_{\mathrm{f}}(\mathcal{H}\otimes\mathcal{K})$.
By the properties of the complementary relation this means
validity of inequality (\ref{super-add}) for the channels
$\widehat{\Phi}$ and $\widehat{\Psi}$ and arbitrary state
$\omega\in\mathfrak{S}_{\mathrm{f}}(\mathcal{H}\otimes\mathcal{K})$.
By the part $\mathrm{(iii)}\Rightarrow\mathrm{(ii)}$ of theorem 2
in \cite{Sh-2} this implies validity of inequality (\ref{sub-add})
for the channels $\widehat{\Phi}$ and $\widehat{\Psi}$ and
arbitrary state
$\omega\in\mathfrak{S}_{\mathrm{f}}(\mathcal{H}\otimes\mathcal{K})$.
By proposition 6 in \cite{Sh-2} the last property implies validity
of inequality (\ref{sub-add}) for the channels $\widehat{\Phi}$
and $\widehat{\Psi}$ and arbitrary state
$\omega\in\mathfrak{S}(\mathcal{H}\otimes\mathcal{K})$, which
means the additivity of the Holevo capacity for the channels
$\widehat{\Phi}$ and $\widehat{\Psi}$ with arbitrary constraints.
$\square$

Propositions \ref{hf-p-3} and \ref{hf-p-4} imply the following
result.

\begin{corollary}\label{hf-c-2}
\textit{Let $\Phi$ be a channel such that its complementary
channel $\widehat{\Phi}$ belongs to one of the classes considered
in proposition \ref{hf-p-3} and $\Psi$ be an arbitrary channel.
Then the superadditivity of the \textup{CCoOE} holds for the
channels $\Phi$ and $\Psi$.}

\textit{If
$H_{\Phi}(\rho)<+\infty,\forall\rho\in\mathrm{ext}\mathfrak{S}(\mathcal{H})$
and
$H_{\Psi}(\sigma)<+\infty,\forall\sigma\in\mathrm{ext}\mathfrak{S}(\mathcal{K})$
then the additivity of the Holevo capacity  holds for the channels
$\Phi$ and $\Psi$ with arbitrary constraints.}
\end{corollary}

By the fist part of corollary \ref{hf-c-2} the superadditivity of
the \textup{CCoOE} holds for arbitrary channel $\Psi$ and the
channel $\Phi_{p}^{a}$ considered in the example in section 3 with
arbitrary probability density function $p(t)$ and $a\leq+\infty$.
Indeed, by using lemma \ref{isomorphism} in Appendix 5.2 it is
easy to see that
$$
\widehat{\Phi}_{p}^{+\infty}(\rho)=\int_{-\infty}^{+\infty}
V_{t}|\varphi_{p}\rangle\langle\varphi_{p}|(V_{t})^{*}\mathrm{Tr}(\rho
M(dt)),\quad\forall\rho\in\mathfrak{S}(\mathcal{H}_{+\infty}),
$$
where $\varphi_{p}(x)=\sqrt{p(x)}$, $V_{t}$ - the shift operator
and $M(\cdot)$ - the projector valued measure of the operator of
multiplication on $x$ in
$\mathcal{H}_{+\infty}=L^{x}_{2}(\mathbb{R})$. By theorem 2 in
\cite{H-Sh-W} the channel $\widehat{\Phi}_{p}^{+\infty}$ is
entanglement-breaking. Since the channel $\Phi_{p}^{a}$ with
$a<+\infty$ can be considered as a subchannel of the channel
$\Phi_{p}^{+\infty}$ corresponding to the subspace
$L^{x}_{2}([-a,a])\subset L^{x}_{2}(\mathbb{R})$ the channel
$\widehat{\Phi}_{p}^{a}$ is entanglement-breaking as well.

By the second part of corollary \ref{hf-c-2} if $a<+\infty$ and
the function $p(t)$ satisfies the conditions assumed in proving
continuity and boundedness of the function
$\hat{H}_{\Phi_{p}^{a}}$ in the example in section 3 then the
additivity of the Holevo capacity holds for the channels
$\Phi_{p}^{a}$ and $\Psi$ with arbitrary constraints, where $\Psi$
is an arbitrary channel such that $H_{\Psi}(\sigma)<+\infty$ for
all $\sigma$ in $\mathrm{ext}\mathfrak{S}(\mathcal{K})$, in
particular, $\Psi=\Phi_{p}^{a}$.

\section{Appendices}

\subsection{Continuity condition for entropy}

For many application it is necessary to have sufficient condition
of continuity of the entropy on some subset of states. There exist
several results of this type, in particular, Simon's convergence
theorems for the entropy \cite{Simon}. The important and widely
used continuity condition is the following \cite{W}: \textit{if
$H$ is a positive unbounded operator in separable Hilbert space
$\mathcal{H}$ such that $\mathrm{Tr}\exp(-\lambda H)<+\infty$ for
all $\lambda>0$ then the entropy is continuous on the set
$\mathcal{K}_{H,h}=\{\rho\in\mathfrak{S}(\mathcal{H})
\,|\,\mathrm{Tr}H\rho\leq h\}$} for any $h>0$.

The feature of this (and some other conditions) consists in
compactness and convexity of the set, on which continuity of the
entropy is established\footnote{Compactness of the set
$\mathcal{K}_{H,h}$ is proved in \cite{H-c-w-c}.}. Thus by using
such conditions we can not prove boundedness and continuity of the
entropy on noncompact or nonconvex sets such that the entropy is
not continuous on their convex closure (note that by corollary 5
in \cite{Sh-4} boundedness of the entropy on a convex set implies
relative compactness of this set). A necessity to deal with such
sets arises when we want to prove  boundedness  and continuity of
the output entropy of a quantum channel on the set of all pure
states if the output entropy of this channel is not bounded on the
whole state space. This problem can be solved by using the
following simple observation.

Let $\mathrm{aut}\mathfrak{S}(\mathcal{H})$ be the set of all
automorphism of the state space $\mathfrak{S}(\mathcal{H})$.

\begin{property}\label{hf-cont-cond}
\textit{Let $\mathcal{K}$ be a compact subset of
$\mathfrak{S}(\mathcal{H})$, on which the entropy is continuous.
Then the entropy is bounded and continuous on the set
$$
\widehat{\mathcal{K}}=\bigcup_{\alpha\in\mathrm{aut}\mathfrak{S}(\mathcal{H})}\alpha(\mathcal{K}).
$$}
\end{property}
It is clear that the entropy is continuous on the "translation"
$\alpha(\mathcal{K})$ of the set $\mathcal{K}$ by given
$\alpha\in\mathrm{aut}\mathfrak{S}(\mathcal{H})$. Compactness of
the set $\mathcal{K}$ implies, loosely speaking, that the entropy
continuously changes under transition between different
translations.

\textbf{Proof.} Boundedness of the entropy is obvious. Suppose the
entropy is not continuous on the set $\widehat{\mathcal{K}}$. Then
there exists a sequence $\{\rho_{n}\}$ of states  in
$\widehat{\mathcal{K}}$ converging to a state $\rho_{0}$ in
$\widehat{\mathcal{K}}$ such that
$\lim_{n\rightarrow+\infty}H(\rho_{n})>H(\rho_{0})$. By definition
of the set $\widehat{\mathcal{K}}$ there exits a sequence
$\{\alpha_{n}\}\subset\mathrm{aut}\mathfrak{S}(\mathcal{H})$ such
that $\sigma_{n}=\alpha_{n}(\rho_{n})\in\mathcal{K}$ for all $n$.
Since the set $\mathcal{K}$ is compact there exists a subsequence
$\{\sigma_{n_{k}}\}$ of the sequence $\{\sigma_{n}\}$ converging
to some state $\sigma_{0}$. Since the states $\rho_{n_{k}}$ and
$\sigma_{n_{k}}=\alpha_{n_{k}}(\rho_{n_{k}})$ are isomorphic for
all $k$ they have the same entropy. Moreover it follows that the
limit states $\sigma_{0}$ and $\rho_{0}$ of the sequences
$\{\rho_{n_{k}}\}$ and $\{\sigma_{n_{k}}\}$ are isomorphic and
hence have the same entropy. By the assumption of continuity of
the entropy on the set $\mathcal{K}$ we have
$$
\lim_{k\rightarrow+\infty}H(\rho_{n_{k}})=
\lim_{k\rightarrow+\infty}H(\sigma_{n_{k}})=H(\sigma_{0})=H(\rho_{0})
$$
contradicting to the definition of the sequence
$\{\rho_{n}\}$.$\square$

For applications it is convenient to reformulate proposition
\ref{hf-cont-cond} as follows.
\begin{corollary}\label{hf-cont-cond-1}
\textit{The entropy is continuous and bounded on a subset
$\mathcal{A}$ of $\mathfrak{S}(\mathcal{H})$ if there exist
separable Hilbert space $\mathcal{K}$ and set
$\{\alpha_{\rho}\}_{\rho\in\mathcal{A}}$ of isomorphisms from
$\mathfrak{S}(\mathcal{H})$ onto $\mathfrak{S}(\mathcal{K})$ such
that $\alpha_{\rho}(\rho)$ lies for each $\rho$ in some compact
subset $\mathcal{B}$ of $\mathfrak{S}(\mathcal{K})$, on which the
entropy is continuous.}
\end{corollary}

Proposition \ref{hf-cont-cond} and the continuity condition for
the entropy stated before imply the following observation. Let
$\mathfrak{U}(\mathcal{H})$ be the group of all unitaries in
$\mathcal{H}$.

\begin{corollary}\label{hf-cont-cond-2}
\textit{Let $H$ be a positive unbounded operator in $\mathcal{H}$
such that $\mathrm{Tr}\exp(-\lambda H)<+\infty$ for all
$\lambda>0$. Then the entropy is continuous on the set
$$
\widehat{\mathcal{K}}_{H,h}=\{\rho\in\mathfrak{S}(\mathcal{H})\,|\,\inf_{U\in\mathfrak{U}(\mathcal{H})}\mathrm{Tr}(HU\rho
U^{*})\leq h\},\quad \forall h>0.
$$}
\end{corollary}

By using corollary \ref{hf-cont-cond-2} it is easy to obtain all
Simon's convergence theorems for entropy \cite{Simon}.

\subsection{The proof of the properties of the channel $\Phi_{p}^{a}$}

In this subsection the properties of the channel $\Phi_{p}^{a}$
described in the example in section 3 are proved. We will show
that the sufficient condition in proposition \ref{hf-p-2} of
boundedness and continuity of the function $\rho\mapsto
H(\Phi_{p}^{a}(\rho))$ on the set
$\mathrm{ext}\mathfrak{S}(\mathcal{H})$ holds for the channel
$\Phi_{p}^{a}$ under the assumed conditions.

Let $\{U_{t}\}_{t\in\mathbb{R}}$ and $\{V_{t}\}_{t\in\mathbb{R}}$
are the unitary groups in the Hilbert space
$\mathcal{L}_{2}(\mathbb{R})$ defined by the expressions
$$
(U_{t}\varphi)(x)=\exp(-\mathrm{i}tx)\varphi(x)\quad\mathrm{and}\quad(V_{t}\varphi)(x)=\varphi(x-t),\quad\forall
\varphi\in\mathcal{L}_{2}(\mathbb{R}).
$$

\begin{lemma}\label{isomorphism}
\textit{The states
$$
\int_{-\infty}^{+\infty}
U_{t}|\psi\rangle\langle\psi|U_{t}^{*}|\widehat{\varphi}(t)|^{2}dt\quad
and\quad \int_{-\infty}^{+\infty}
V_{t}|\varphi\rangle\langle\varphi|V_{t}^{*}|\psi(t)|^{2}dt
$$
are isomorphic for arbitrary $\varphi$ and $\psi$ in
$\mathcal{L}_{2}(\mathbb{R})$ with $\|\varphi\|=1$ and
$\|\psi\|=1$, where $\widehat{\varphi}$ is the Fourier
transformation of $\varphi$.}
\end{lemma}

\textbf{Proof.} By direct calculation it is possible to show that
the above states coincide with the partial traces
$\mathrm{Tr}_{\mathcal{L}^{x}_{2}}|\theta\rangle\langle\theta|$
and
$\mathrm{Tr}_{\mathcal{L}^{y}_{2}}|\theta\rangle\langle\theta|$ of
the pure state $|\theta\rangle\langle\theta|$ in
$\mathfrak{S}(\mathcal{L}^{x}_{2}\otimes \mathcal{L}^{y}_{2})$,
where $\theta(x,y)=\varphi(x-y)\psi(y)$. $\square$

Suppose first that the function $p(t)$ has finite support. Then
there exist $\lambda>0$ and $b>0$ such that $\lambda p(t)\leq
\displaystyle\frac{\sin^{2}bt}{\pi b t^{2}}$ for all
$t\in\mathbb{R}$ and hence $\lambda\rho_{\psi}\leq\sigma_{\psi}$,
where
$$
\rho_{\psi}=\int_{-\infty}^{+\infty}
U_{t}|\psi\rangle\langle\psi|U_{t}^{*}p(t)dt\quad\textup{and}\quad\sigma_{\psi}=\int_{-\infty}^{+\infty}
U_{t}|\psi\rangle\langle\psi|U_{t}^{*}\frac{\sin^{2}bt}{\pi b
t^{2}}dt.
$$
for arbitrary  $\psi$ in $\mathcal{L}_{2}([-a,a])=\mathcal{H}_{a}$
with $\|\psi\|=1$.

By lemma \ref{isomorphism} for each $\psi\in\mathcal{H}_{a}$ with
$\|\psi\|=1$ there exists unitary transformation
$W_{\psi}:\mathcal{H}_{a}\mapsto\mathcal{K}=\mathcal{L}_{2}(\mathbb{R})$
such that
$$
W_{\psi}\sigma_{\psi}W_{\psi}^{*}=\tilde{\sigma}_{\psi}=\int_{-\infty}^{+\infty}
V_{t}|\varphi_{b}\rangle\langle\varphi_{b}|V_{t}^{*}|\psi(t)|^{2}dt,
$$
where $\varphi_{b}(x)=\left\{
       \begin{array}{ll}
       1, & x\in[-b,b]\\
       0, & x\in\mathbb{R}\setminus[-b,b].
       \end{array}\right.$

Let $c=a+b$ and $\tilde{V}^{c}_{t}$ be the operator of cyclic
shift on $t$ in $\mathcal{L}_{2}([-c,c])$ considered as a subspace
of $\mathcal{L}_{2}(\mathbb{R})$ consisting of functions supported
by $[-c,c]$.

Note that
$\tilde{V}^{c}_{t}|\varphi_{b}\rangle=V_{t}|\varphi_{b}\rangle$
for all $t$ in $[-a,a]$. Since $\mathrm{supp}\psi\subseteq[-a,a]$
we have
\begin{equation}\label{hf-exp-1}
\tilde{\sigma}_{\psi}=\int_{-a}^{a}
V_{t}|\varphi_{b}\rangle\langle\varphi_{b}|V_{t}^{*}|\psi(t)|^{2}dt=
\int_{-a}^{a}
\tilde{V}^{c}_{t}|\varphi_{b}\rangle\langle\varphi_{b}|(\tilde{V}^{c}_{t})^{*}|\psi(t)|^{2}dt.
\end{equation}

It is clear that
$\lambda\tilde{\rho}_{\psi}\leq\tilde{\sigma}_{\psi}$, where
$\tilde{\rho}_{\psi}=W_{\psi}\rho_{\psi}W_{\psi}^{*}$.

Let $H=\sum_{k}\log^{2}(|k|+1)|k\rangle\langle k|$ be a positive
operator in $\mathcal{L}_{2}([-c,c])$,  where
$\{|k\rangle\sim(2c)^{1/2}\exp(\mathrm{i}\pi c^{-1}kx)\}$ -
trigonometric basic in $\mathcal{L}_{2}([-c,c])$. Note that
$\mathrm{Tr}\exp(-\lambda H)<+\infty$ for all $\lambda>0$. Since
$\tilde{V}^{c}_{t}=\sum_{k=-\infty}^{k=+\infty}\exp(-\mathrm{i}\pi
c^{-1}kt)|k\rangle\langle k|$ we obtain from (\ref{hf-exp-1}) that
$$
\begin{array}{c}
\displaystyle\mathrm{Tr}\tilde{\sigma}_{\psi}H=\sum_{k=-\infty}^{k=+\infty}\log^{2}(|k|+1)\langle
k|\tilde{\sigma}_{\psi}|k\rangle=\sum_{k=-\infty}^{k=+\infty}\log^{2}(|k|+1)|\langle
\varphi_{b}|k\rangle|^{2}\\\\\displaystyle=\sum_{k=-\infty}^{k=+\infty}\log^{2}(|k|+1)|\frac{c\sin^{2}(\pi
c^{-1}kb)}{b\pi^{2}k^{2}}=h<+\infty,\quad\forall\psi,
\end{array}
$$
and hence
$\mathrm{Tr}\tilde{\rho}_{\psi}H\leq\lambda^{-1}\mathrm{Tr}\tilde{\sigma}_{\psi}H=\lambda^{-1}h<+\infty,\;\forall\psi$.

Thus the sufficient condition in proposition \ref{hf-p-2} is
fulfilled with $\mathcal{K}=\mathcal{L}_{2}(\mathbb{R})$,
$\alpha_{\psi}=W_{\psi}(\cdot)W_{\psi}^{*}$ and
$\mathcal{A}=\{\rho\in
\mathfrak{S}(\mathcal{L}_{2}([-c,c]))\subset\mathfrak{S}(\mathcal{K})\,|\,\mathrm{Tr}\rho
H\leq \lambda^{-1}h\}$ (the set $\mathcal{A}$ has the required
properties by the observation in \cite{H-Sh-2}). By this condition
the function $\psi\mapsto
H_{\Phi_{p}^{a}}(|\psi\rangle\langle\psi|)=H(\rho_{\psi})$ is
bounded and continuous.

Note that this implies in particular that
\begin{equation}\label{ent-finite}
H(\omega(\psi))\leq C<+\infty\;\textup{for}\;\textup{all}\;\psi,\;
\textup{where}\;\,
\omega(\psi)=\int_{0}^{1}U_{t}|\psi\rangle\langle\psi|U_{t}^{*}dt.
\end{equation}

Let $p(t)$ be a bounded density function such that
$|\int_{-\infty}^{+\infty}p(t)\log p(t)dt|<+\infty$ and for
sufficiently large $d$ the function $p(t)$ is monotonous on
$(-\infty,-d)$ and on $(d, +\infty)$. In what follows we will
assume that $d$ is so large that the both functions $p(t)$ and
$p(t)\log p(t)$ are monotonous on $(-\infty,-d+1)$ and on $(d-1,
+\infty)$.

Let
$\rho^{d}_{\psi}=\left(\int_{-d}^{d}p(t)dt\right)^{-1}\int_{-d}^{d}
U_{t}|\psi\rangle\langle\psi|U_{t}^{*}p(t)dt$. Below we will prove
that for all sufficiently large $d$ the following inequality holds
\begin{equation}\label{hf-tails}
\begin{array}{c}
\displaystyle|H(\rho_{\psi})-H(\rho^{d}_{\psi})|\leq
\alpha(d)H(\rho^{d}_{\psi})+h_{2}(\alpha(d))+
\alpha(d-1)C+\beta(d-1)\\\\ \displaystyle\leq \frac{\alpha(d)}{1-
\alpha(d)}H(\rho_{\psi})+h_{2}(\alpha(d))+\alpha(d-1)C+\beta(d-1),
\end{array}
\end{equation}
where $C$ is defined in (\ref{ent-finite}),
$\alpha(d)=\int_{|t|>d}p(t)dt$ and $
\beta(d)=|\int_{|t|>d}p(t)\log p(t)dt|$. The first inequality in
(\ref{hf-tails}) with fixed $d$ and the above observation
(concerning the function $p(t)$ with finite support) imply uniform
boundedness of the function $\psi\mapsto
H_{\Phi_{p}^{a}}(|\psi\rangle\langle\psi|)=H(\rho_{\psi})$. The
second inequality in (\ref{hf-tails}) with $d$ increasing to
$+\infty$ and the above observation imply continuity of the
function $\psi\mapsto
H_{\Phi_{p}^{a}}(|\psi\rangle\langle\psi|)=H(\rho_{\psi})$.

To prove (\ref{hf-tails}) consider convex decomposition
$\rho_{\psi}=(1- \alpha(d))\rho_{\psi}^{d}+
\alpha(d)\sigma_{\psi}^{d}$, where $\sigma^{d}_{\psi}=
\alpha^{-1}(d)\int_{|t|>d}
U_{t}|\psi\rangle\langle\psi|U_{t}^{*}p(t)dt$. By the property of
the entropy the above decomposition implies
\begin{equation}\label{concavity}
\begin{array}{c}
\displaystyle(1- \alpha(d))H(\rho^{d}_{\psi})+
\alpha(d)H(\sigma_{\psi}^{d})\\\\ \displaystyle\leq
H(\rho_{\psi})\leq\\\\ \displaystyle (1-
\alpha(d))H(\rho^{d}_{\psi})+ \alpha(d)H(\sigma_{\psi}^{d})+h_{2}(
\alpha(d))
\end{array}
\end{equation}

and hence
\begin{equation}\label{hf-ineq-1}
 |H(\rho_{\psi})-H(\rho^{d}_{\psi})|\leq
 \alpha(d)H(\rho^{d}_{\psi})+ \alpha(d)H(\sigma_{\psi}^{d})+h_{2}( \alpha(d)).
\end{equation}

By using monotonicity of the function $p(t)$ on $(-\infty,-d+1)$
and on $(d-1, +\infty)$ we obtain the following operator
inequality
\begin{equation}\label{hf-l-ineq}
\begin{array}{c}
\displaystyle
\alpha(d)\sigma_{\psi}^{d}=\int_{-\infty}^{-d}U_{t}|\psi\rangle\langle\psi|U_{t}^{*}p(t)dt+
\int_{d}^{+\infty}U_{t}|\psi\rangle\langle\psi|U_{t}^{*}p(t)dt\\\\\displaystyle\leq
\sum_{k=-\infty}^{-1}p(-d+k+1)\int_{-d+k}^{-d+k+1}U_{t}|\psi\rangle\langle\psi|U_{t}^{*}dt
\\\\\displaystyle+\sum_{k=0}^{+\infty}p(d+k)\int_{d+k}^{d+k+1}U_{t}|\psi\rangle\langle\psi|U_{t}^{*}dt=
\gamma(d)\sum_{k=-\infty}^{k=+\infty}\lambda_{k}\omega_{k}(\psi),
\end{array}
\end{equation}
where
$\gamma(d)=\sum_{k=-\infty}^{-1}p(-d+k+1)+\sum_{k=0}^{+\infty}p(d+k)$,
$$
\lambda_{k}=\left\{
   \begin{array}{ll}
    \gamma^{-1}(d)p(-d+k+1) & k<0\\
    \gamma^{-1}(d)p(d+k), & k\geq 0,
    \end{array}\right.
\quad
\omega_{k}(\psi)=\int_{l_{k}}^{l_{k}+1}U_{t}|\psi\rangle\langle\psi|U_{t}^{*}dt,
$$
$l_{k}=-d+k$ if $k<0$ and $l_{k}=d+k$ if $k\geq0$.

Since for each $k$ the state $\omega_{k}(\psi)$ is isomorphic to
the state $\omega(\psi)$ defined in (\ref{ent-finite}) we have
$H(\omega_{k}(\psi))\leq C$. Inequality (\ref{hf-l-ineq}) and the
property of the entropy imply
\begin{equation}\label{hf-est}
 \alpha(d)H(\sigma_{\psi}^{d})\leq\gamma(d)C-\gamma(d)\sum_{k=-\infty}^{k=+\infty}\lambda_{k}\log\lambda_{k}.
\end{equation}

By using monotonicity of the function $p(t)\log p(t)$ on
$(-\infty,-d+1)$ and on $(d-1, +\infty)$ we obtain
\begin{equation}\label{hf-ineq-2}
\begin{array}{c}
\displaystyle-\gamma(d)\sum_{k=-\infty}^{k=+\infty}\lambda_{k}\log\lambda_{k}=\gamma(d)\log\gamma(d)\\\\\displaystyle-
\sum_{k=-\infty}^{-1}p(-d+k+1)\log
p(-d+k+1)-\sum_{k=0}^{+\infty}p(d+k)\log
p(d+k)\\\\\displaystyle\leq
\gamma(d)\log\gamma(d)-\int_{|t|>d-1}p(t)\log
p(t)dt\leq\gamma(d)\log\gamma(d)+\beta(d-1).
\end{array}
\end{equation}

By noting that $\gamma(d)\leq \alpha(d-1)<1$ for all sufficiently
large $d$ (due to monotonicity of the function $p(t)$ on
$(-\infty,-d+1)$ and on $(d-1, +\infty)$ and since
$\alpha(d)\rightarrow 0$ as $d\rightarrow+\infty$) we obtain from
(\ref{hf-ineq-1}), (\ref{hf-est}) and (\ref{hf-ineq-2}) the first
inequality in (\ref{hf-tails}). The second inequality in
(\ref{hf-tails}) follows from the inequality
$(1-\alpha(d))H(\rho^{d}_{\psi})\leq H(\rho_{\psi})$ easily
deduced from (\ref{concavity}).

\textbf{Acknowledgments.} I am grateful to A.S.Holevo for the
permanent help. I would like also to thank J.I.Cirac and M.M.Wolf
for the useful discussion during my visit the Max Plank Institute
for Quantum Optics. The work was partially supported by the
program "Modern problems of theoretical mathematics" of Russian
Academy of Sciences and by Russian Foundation of Fundamental
Research (project N 06-01-00164-a).

\end{document}